\documentclass[english,english]{article}
\usepackage[T1]{fontenc}
\usepackage[latin1]{inputenc}

\makeatletter

\providecommand{\LyX}{L\kern-.1667em\lower.25em\hbox{Y}\kern-.125emX\@}

 \newcommand{\lyxaddress}[1]{
   \par {\raggedright #1 
   \vspace{1.4em}
   \noindent\par}
 }

\usepackage[T1]{fontenc}
\usepackage[latin1]{inputenc}

\makeatletter



\makeatother
\begin{document}

\title{Transport Processes in Metal-Insulator Granular Layers }

\author{Yu.G. Pogorelov \thanks{
Author to whom correspondence should be addressed 
} and J. F. Polido}

\maketitle

\lyxaddress{CFP/Departamento de Física, Faculdade de Ciências, Universidade do
Porto, Rua do Campo Alegre, 687, 4169-007 Porto, Portugal}

\begin{abstract}
The non-equilibrium tunnel transport processes are considered in a
square lattice of metallic nanogranules embedded into insulating host.
Based on a simple model with three possible charging states (\( \pm  \),
or \( 0 \)) of a granule and three kinetic processes (creation or
recombination of a \( \pm  \) pair, and charge translation) between
neighbor granules, the mean-field kinetic theory is developed, which
takes account of the interplay between charging energy and temperature
and between the applied electric field and the Coulomb fields by non-compensated
charge density. The resulting charge and current distributions are
found to depend essentially on the particular conditions in a granular
layer, namely, in a free area (FA) or in contact areas (CA) under
macroscopic metallic contacts. Thus, a steady state dc transport is
only compatible with zero charge density and ohmic resistivity within
FA, but charge accumulation and non-ohmic behavior are \emph{necessary}
for conduction over CA. The approximate analytic solutions are obtained
for characteristic regimes (low or high charge density) of such conduction.
Also non-stationary processes are considered, displaying a peculiar
combination of two strongly different relaxation times. The comparison
is done with the available transport data on similar experimental
systems.

\end{abstract}

\section{Introduction}

\label{int}Granular (including nanogranular) systems are of a considerable
interest for modern technology due to their new physical properties,
like giant magnetoresistance \cite{Berk}, Coulomb blocade \cite{Fert},\cite{Ima}
or high density magnetic memory \cite{Park}, impossible for classical
materials. The systems of our interest here are composed from nanoscopic
(spherical) metallic particles embedded into insulating matrix, and
various existing experimental techniques result in ordered or disordered,
mono- or polydispersed, single- or multilayered structures \cite{Dieny}.
They are of no less interest from the point of view of fundamental
study. In particular, transport phenomena in granular systems reveal
certain characteristics which cannot be obtained neither in classical
flow regime (in metallic, electrolyte, or gas discharge conduction)
nor in hopping regime (in doped semiconductors or in common tunnel
junctions). Their specifics is mainly determined by the drastic difference
between the characteristic time of an individual tunneling event (\( \sim \hbar /\varepsilon _{{\textrm{F}}}\sim 10^{-15} \)
s) and the interval between such events on the same granule (\( \sim e/(jd^{2})\sim 10^{-3} \)
s, at typical current density \( j\sim 10^{-3} \) A/cm\( ^{2} \)
and granule diameter \( d\sim 5 \) nm). Other important moments are
the sizeable Coulomb charging energy \( E_{c}\sim e^{2}/\varepsilon _{eff}d \)
(typically \( \sim 10 \) meV) and the fact that the tunneling rates
across the layer may be even several orders of magnitude slower than
along it. The interplay of all these factors leads to unusual macroscopic
effects, thus, a peculiar slow relaxation of electric charge was recently
discovered in experiments on tunnel conduction through granular layers
and granular films \cite{Kak},\cite{Sch}. For theoretical description
of such processes we use the model of a single layer of identical
spherical particles located in sites of a simple square lattice, with
three possible charging states (\( \pm  \), or \( 0 \)) of a granule
and three kinetic processes (creation or recombination of a \( \pm  \)
pair, and charge translation) between neighbor granules, and even
such simple model reveals a variety of possible dynamical and thermodynamical
regimes, to be presented below.

The detailed formulation of the model, its basic parameters, and its
mean-field continuum version are given in Sec. \ref{char}. Next in
Sec. \ref{mf} we calculate the mean values of occupation numbers
of each charging state for steady state conditions, including the
simplest equilibrium situation (no applied fields), in function of
temperature. The analysis of current density and related kinetic equation
in the out-of-equilibrium case is developed in Sec. \ref{fa}, where
also its simple solution is discussed for the FA part of the system.
The most non-trivial regimes are found for the CA part, as described
in Sec. \ref{ca} for steady state conduction with charge accumulation
and non-ohmic behavior. At least, some non-stationary solutions are
considered in Sec. \ref{tdep}, either obtained straightforwardly
from the time-dependent kinetic equation or estimated from a simplified
equivalent circuit. The general integration scheme for non-linear
differential equation, corresponding to steady state in CA, and the
particular approximations leading to its analytic solutions are dropped
into Appendix.

\section{\label{char}Charging states and kinetic processes }

We consider a system of identical spherical metallic nanogranules
of diameter \( d \), located in sites of simple square lattice of
period \( a \) within a layer of thickness \( b\sim a \) of insulating
host with dielectric constant \( \varepsilon  \) (Fig. 1). In the
charge transfer processes, granules can bear different numbers \( \sigma  \)
of electrons in excess (or deficit) to the constant number of positive
ions, so that the resulting excess charge \( \sigma e \) defines
a certain Coulomb charging energy \( E_{c} \). At low enough temperatures,
which are implied in what follows, the consideration can be limited,
besides the ground neutral state \( \sigma =0 \), only to single
charged states \( \sigma =\pm 1 \). For the latter case, \( E_{c} \)
was given in the classic paper by Sheng and Abeles \cite{ShAb} in
the form \( E_{c}=e^{2}f(s/d)/(\varepsilon d) \), where \( s \)
is the mean spacing between granules and the dimensionless function
\( f(z) \) was estimated for a 3D granular array as \( f(z)=1/(1+1/2z) \).
Otherwise, the mutual dielectric response of 3D insulating host (\( \varepsilon _{h}=\varepsilon  \))
with volume fraction \( f<1 \) of metallic particles (\( \varepsilon _{m}\rightarrow \infty  \))
can be characterized by an effective value \( \varepsilon _{eff}=\varepsilon /(1-f) \).
For the planar lattice of granules, this effective constant can be
estimated, summing the own energy \( e^{2}/\varepsilon d \) of a
charged granule at site 0 with the energy of its interaction with
electric dipolar moments \( \approx (e/\varepsilon _{eff})(d/2n)^{3}\mathbf{n} \),
induced in each granule at site \( \mathbf{n}=a(n_{1},n_{2}) \):\begin{equation}
\label{eq1}
E_{c}=\frac{e^{2}}{d}\left[ \frac{1}{\varepsilon }-\frac{\alpha }{\varepsilon _{eff}}\left( \frac{d}{a}\right) ^{4}\right] =\frac{e^{2}}{\varepsilon _{eff}d}.
\end{equation}
 Here the constant \( \alpha =(a^{4}/8)\sum _{\mathbf{n}\neq 0}n^{-4}\approx 0.92 \),
and the resulting \( \varepsilon _{eff}=\varepsilon (1+\alpha (d/a)^{4}) \).
However, Eq. \ref{eq1} may underestimate considerably the most important
screening from nearest neighbor granules at \( d\rightarrow a \),
and in what follows we generally characterize the composite of insulating
matrix and metallic granules by a certain \( \varepsilon _{eff}=e^{2}/dE_{c}\gg \varepsilon  \).

Following the approach proposed earlier \cite{Kak}, we classify the
microscopic states of our system, attributing the charging variable
\( \sigma _{\mathbf{n}} \) with values \( \pm 1 \) or \( 0 \) to
each site \( \mathbf{n} \) and then considering three types of kinetic
processes between two neighbor granules \( \mathbf{n} \) and \( \mathbf{n}+{\Delta } \)
(Fig. 2):

1) electron hopping from neutral \( \mathbf{n} \) to neutral \( \mathbf{n}+{\Delta } \),
creating a pair of oppositely charged granules: \( (\sigma _{\mathbf{n}}=0,\sigma _{\mathbf{n}+{\Delta }}=0)\rightarrow (\sigma _{\mathbf{n}}=-1,\sigma _{\mathbf{n}+{\Delta }}=1) \),
it is only this process that was included in Sheng and Abeles' theory;

2) hopping of an extra electron (hole) from \( \mathbf{n} \) to neutral
\( \mathbf{n}+{\Delta } \), that is charge transfer: \( (\sigma _{\mathbf{n}}=1,\sigma _{\mathbf{n}+{\Delta }}=0)\rightarrow (\sigma _{\mathbf{n}}=0,\sigma _{\mathbf{n}+{\Delta }}=1) \),
or \( (\sigma _{\mathbf{n}}=-1,\sigma _{\mathbf{n}+{\Delta }}=0)\rightarrow (\sigma _{\mathbf{n}}=0,\sigma _{\mathbf{n}+{\Delta }}=-1) \);

3) an inverse process to 1), that is recombination of a pair: \( (\sigma _{\mathbf{n}}=1,\sigma _{\mathbf{n}+{\Delta }}=-1)\rightarrow (\sigma _{\mathbf{n}}=0,\sigma _{\mathbf{n}+{\Delta }}=0) \).

Note that all the processes 1) to 3) are conserving the total system
charge \( Q=\sum _{\mathbf{n}}\sigma _{\mathbf{n}} \), hence the
possibility for charge accumulation or relaxation only appears due
to current leads. A typical configuration for current-in-plane (CIP)
tunneling conduction includes two macroscopic metallic electrodes
on top of the granular layer, forming contact areas (CA) where the
current is being distributed from the electrodes into granules through
an insulating spacer of thickness \( b^{\prime } \), and a free area
(FA) where the current propagates between the contacts (Fig. 3). To
begin with, let us consider a simpler case of FA while the specific
analysis for CA with an account for screening effects by metallic
contacts will be given later in Sec. \ref{ca}.

The respective transition rates \( q_{i} \) for the processes 1)
to 3) are determined either by the instantaneous charging states of
two relevant granules and by the local electric field \( \mathbf{F}_{\mathbf{n}} \)
and temperature \( T \), accordingly to the expressions:\begin{equation}
\label{eq2}
q^{\left( 1\right) }_{\mathbf{n},{\Delta }}=\left( 1-\sigma _{\mathbf{n}}^{2}\right) \left( 1-\sigma _{\mathbf{n}+{\Delta }}^{2}\right) \varphi \left( e{\mathbf{F}_{\mathbf{n}}\cdot \Delta }+E_{c}\right) ,
\end{equation}
 \[
q^{\left( 2\right) }_{\mathbf{n},{\Delta }}=\sigma _{\mathbf{n}}^{2}\left( 1-\sigma _{\mathbf{n}+{\Delta }}^{2}\right) \varphi \left( -e{\mathbf{F}_{\mathbf{n}}\cdot \Delta }\right) ,\]
 \[
q^{\left( 3\right) }_{\mathbf{n},{\Delta }}=\frac{\sigma _{\mathbf{n}}\sigma _{\mathbf{n}+{\Delta }}\left( \sigma _{\mathbf{n}}\sigma _{\mathbf{n}+{\Delta }}-1\right) }{2}\varphi \left( e\sigma _{\mathbf{n}+{\Delta }}{\mathbf{F}_{\mathbf{n}}\cdot \Delta }-E_{c}\right) .\]
 Here the charging energy is taken positive, \( E_{c} \), for pair
creation, zero for transport, and negative, \( -E_{c} \), for recombination.
The function \( \varphi \left( E\right) =\omega N_{{\textrm{F}}}E/({\textrm{e}}^{\beta E}-1) \)
expresses the total probability, at given \( \beta =1/k_{{\textrm{B}}}T \),
for electron transition between granules with Fermi density of states
\( N_{{\textrm{F}}} \) and Fermi levels differing by \( E \). The
hopping frequency \( \omega =\omega _{a}{\textrm{e}}^{-2\chi s} \)
is expressed through the {}``attempt{}'' frequency \( \omega _{a}\sim \varepsilon _{{\textrm{F}}}/\hbar  \),
the inverse tunneling length \( \chi  \) (typically \( \sim 10 \)
nm\( ^{-1} \)), and the intergranule spacing \( s=a-d \). Local
electric field \( \mathbf{F}_{\mathbf{n}} \) on \( \mathbf{n} \)th
site consists of the external field \( \mathbf{F}_{{ext}} \) and
the Coulomb field \( {\mathbf{F}}_{\mathbf{n}}^{Coul} \) due to all
other charged granules (from FA):\begin{equation}
\label{eq3}
{\mathbf{F}}_{\mathbf{n}}^{Coul}=\frac{e}{\varepsilon _{eff}}\sum _{\mathbf{n}^{\prime }\neq \mathbf{n}}\sigma _{\mathbf{n}^{\prime }}\frac{\mathbf{n}^{\prime }-\mathbf{n}}{\left| \mathbf{n}^{\prime }-\mathbf{n}\right| ^{3}}.
\end{equation}

A suitable approximation for such a system is achieved with passing
from discrete-valued functions \( \sigma _{\mathbf{n}} \) of discrete
argument \( {\mathbf{n}}=a(n_{1},n_{2}) \) to their continuous-valued
mean-field (MF) equivalents \( \sigma ({\mathbf{r}})=\left\langle \sigma _{\mathbf{n}}\right\rangle _{\mathbf{r}} \)
(mean charge density) and \( \rho ({\mathbf{r}})=\left\langle \sigma ^{2}_{\mathbf{n}}\right\rangle _{\mathbf{r}} \)
(mean charge carrier density), obtained by averaging over a wide enough
(that is, great compared to the lattice period but small compared
to the size of entire system or its parts) area around \( \mathbf{r} \),
the latter being \emph{any} point in the plane. This also implies
passing to smooth local field:\begin{equation}
\label{eq4}
{\mathbf{F}(\mathbf{r})}={\mathbf{F}}_{ext}+\frac{e}{\varepsilon _{eff}a^{2}}\int \sigma ({\mathbf{r}^{\prime }})\frac{\mathbf{r}^{\prime }-\mathbf{r}}{\left| \mathbf{r}^{\prime }-\mathbf{r}\right| ^{3}}d\mathbf{r}^{\prime }.
\end{equation}
 and to averaged transition rates \( q^{\left( i\right) }({\mathbf{r},\Delta })=\left\langle q^{\left( i\right) }_{\mathbf{n},{\Delta }}\right\rangle _{\mathbf{r}} \)
and \( p^{\left( i\right) }({\mathbf{r},\Delta })=\left\langle \sigma _{\mathbf{n}}q^{\left( i\right) }_{\mathbf{n},{\Delta }}\right\rangle _{\mathbf{r}} \).
These rates just define the temporal evolution of mean densities:\begin{equation}
\label{eq5}
\dot{\sigma }({\mathbf{r}})=\sum _{{\Delta }}\left[ q^{\left( 1\right) }({\mathbf{r},\Delta })-q^{\left( 1\right) }({\mathbf{r}+\Delta ,-\Delta })-p^{\left( 2\right) }({\mathbf{r},\Delta })+\right. 
\end{equation}
 \[
\left. +p^{\left( 2\right) }({\mathbf{r}+\Delta ,-\Delta })-p^{\left( 3\right) }({\mathbf{r}},{\Delta })\right] ,\]
\begin{equation}
\label{eq5a}
\dot{\rho }({\mathbf{r}})=\sum _{{\Delta }}\left[ q^{\left( 1\right) }({\mathbf{r},\Delta })+q^{\left( 1\right) }({\mathbf{r}+\Delta ,-\Delta })-q^{\left( 2\right) }({\mathbf{r},\Delta })+\right. 
\end{equation}
 \[
\left. +q^{\left( 2\right) }({\mathbf{r}+\Delta ,-\Delta })-q^{\left( 3\right) }({\mathbf{r}},{\Delta })\right] .\]

The set of Eqs. \ref{eq2}-\ref{eq5a} is complete to provide a continuous
description of the considered system, once a proper averaging procedure
is established.

\section{\label{mf}Mean-field densities in equilibrium}

We perform the above defined averages in the simplest assumption of
no correlations between different sites: \( \left\langle f_{\mathbf{n}}g_{\mathbf{n}^{\prime }}\right\rangle =\left\langle f_{\mathbf{n}}\right\rangle \left\langle g_{\mathbf{n}^{\prime }}\right\rangle  \),
\( \mathbf{n}^{\prime }\neq \mathbf{n}, \) and using the evident
rules:\( \left\langle \sigma ^{2k+1}_{\mathbf{n}}\right\rangle _{\mathbf{r}}=\sigma ({\mathbf{r}}) \),
\( \left\langle \sigma ^{2k}_{\mathbf{n}}\right\rangle _{\mathbf{r}}=\rho ({\mathbf{r}}) \).
The resulting averaged rates are:\begin{equation}
\label{eq6}
q^{\left( 1\right) }({\mathbf{r},\Delta })=\sigma ^{0}({\mathbf{r}})\sigma ^{0}({\mathbf{r}+\Delta })\varphi \left[ e{\mathbf{F}(\mathbf{r})\cdot \Delta }+E_{c}\right] ,
\end{equation}
 \[
q^{\left( 2\right) }({\mathbf{r},\Delta })=\sigma ^{0}({\mathbf{r}+\Delta })\left\{ \sigma ^{+}({\mathbf{r}})\varphi \left[ -e{\mathbf{F}(\mathbf{r})\cdot \Delta }\right] +\sigma ^{-}({\mathbf{r}})\varphi \left[ e{\mathbf{F}(\mathbf{r})\cdot \Delta }\right] \right\} ,\]
 \[
p^{\left( 2\right) }({\mathbf{r},\Delta })=\sigma ^{0}({\mathbf{r}+\Delta })\left\{ \sigma ^{+}({\mathbf{r}})\varphi \left[ -e{\mathbf{F}(\mathbf{r})\cdot \Delta }\right] -\sigma ^{-}({\mathbf{r}})\varphi \left[ e{\mathbf{F}(\mathbf{r})\cdot \Delta }\right] \right\} ,\]
 \[
q^{\left( 3\right) }({\mathbf{r},\Delta })=\left\{ \sigma ^{-}({\mathbf{r}})\sigma ^{+}({\mathbf{r}+\Delta })\varphi \left[ e{\mathbf{F}(\mathbf{r})\cdot \Delta }-E_{c}\right] +\right. \]
\[
\left. +\sigma ^{+}({\mathbf{r}})\sigma ^{-}({\mathbf{r}+\Delta })\varphi \left[ -e{\mathbf{F}(\mathbf{r})\cdot \Delta }-E_{c}\right] \right\} ,\]
 \[
p^{\left( 3\right) }({\mathbf{r},\Delta })=\left\{ \sigma ^{-}({\mathbf{r}})\sigma ^{+}({\mathbf{r}+\Delta })\varphi \left[ e{\mathbf{F}(\mathbf{r})\cdot \Delta }-E_{c}\right] -\right. \]
\[
\left. -\sigma ^{+}({\mathbf{r}})\sigma ^{-}({\mathbf{r}+\Delta })\varphi \left[ -e{\mathbf{F}(\mathbf{r})\cdot \Delta }-E_{c}\right] \right\} ,\]
 where the mean occupation numbers for each charging state \( \sigma ^{\pm }({\mathbf{r}})=[\rho ({\mathbf{r}})\pm \sigma ({\mathbf{r}})]/2 \)
and \( \sigma ^{0}({\mathbf{r}})=1-\rho ({\mathbf{r}}) \) satisfy
the normalization condition: \( \sum _{i}\sigma ^{i}({\mathbf{r}})=1 \).

In a similar way to Eq. \ref{eq5}, we express the vector of average
current density \( \mathbf{j}\left( n\right)  \) at \( \mathbf{n} \)th
site:\begin{equation}
\label{eq7}
{\mathbf{j}\left( n\right) }=\frac{e}{a^{2}b}\sum _{{\Delta }}{\Delta }\left[ -q^{\left( 1\right) }({\mathbf{n},\Delta })+q^{\left( 1\right) }({\mathbf{n}+\Delta ,-\Delta })+p^{\left( 2\right) }({\mathbf{n},\Delta })-\right. 
\end{equation}
 \[
\left. -p^{\left( 2\right) }({\mathbf{n}+\Delta ,-\Delta })+p^{\left( 3\right) }({\mathbf{n},\Delta })\right] ,\]
 then its MF extension \( \mathbf{j}(r) \) is obtained by simple
replacing \( \mathbf{n} \) by \( \mathbf{r} \) in the arguments
of \( q^{\left( i\right) } \) and \( p^{\left( i\right) } \). Expanding
these continuous functions in powers of \( |{\Delta }|=a \), we conclude
that Eq. \ref{eq5} reduces to usual continuity equation:\begin{equation}
\label{eq8}
\dot{\sigma }\left( {\mathbf{r}}\right) =-\frac{a^{2}b}{e}{\nabla }_{2}\cdot {\mathbf{j}\left( r\right) },
\end{equation}
 with the 2D nabla: \( {\nabla }_{2}=(\partial /\partial x,\partial /\partial y) \).

We begin the analysis of Eqs. \ref{eq5}-\ref{eq8} from the simplest
situation of thermal equilibrium in absence of electric field, \( \mathbf{F}(r)\equiv 0 \),
then Eq. \ref{eq5} turns into evident identity: \( \sigma ({\mathbf{r}})\equiv 0 \),
that means zero charge density, and Eq. \ref{eq7} yields in zero
current density: \( \mathbf{j}(r)\equiv 0 \), while Eq. \ref{eq5a}
provides a finite and constant value of charge carrier density:\begin{equation}
\label{eq9}
\rho ({\mathbf{r}})\equiv \rho _{0}=\frac{2}{2+\exp \left( \beta E_{c}/2\right) }.
\end{equation}

At low temperatures, \( \beta E_{c}\gg 1 \), this value is exponentially
small: \( \rho _{0}\approx 2\exp (-\beta E_{c}/2) \), and for high
temperatures, \( \beta E_{c}\ll 1 \), it tends as \( \rho _{0}\approx \rho _{\infty }-\beta E_{c}/9 \)
to the limit \( \rho _{\infty }=2/3 \), corresponding to equipartition
between all three fractions \( \sigma ^{i} \) (Fig. 4).

In presence of electric fields \( \mathbf{F}(r)\neq 0 \), the local
equilibrium should be perturbed and the system can generate current
and possibly accumulate charge. Our further analysis is aimed to demonstrate
that the latter process is impossible in FA, but, supposing for a
moment a non-zero charge density \( \sigma ({\mathbf{r}}) \), we
obtain from Eq. \ref{eq5a} a more general local relation between
\( \sigma ({\mathbf{r}}) \) and \( \rho ({\mathbf{r}}) \):\begin{equation}
\label{eq10}
\sigma ^{2}({\mathbf{r}})=\frac{\left[ \rho ({\mathbf{r}})-\rho _{0}\right] \left[ \rho ({\mathbf{r}})+\rho _{0}-2\rho _{0}\rho ({\mathbf{r}})\right] }{1-\rho _{0}},
\end{equation}
 describing the increase of charge density with going away from equlibrium
(Fig. 5). Now we are in position to pass to the out-of-equilibrium
situations, beginning from a simpler case of dc current flowing through
FA.

\section{\label{fa}Steady state conduction in FA}

In presence of (generally non-uniform) fields \( \mathbf{F}(r) \)
and densities \( \sigma ({\mathbf{r}}) \), \( \rho ({\mathbf{r}}) \),
we expand Eq. \ref{eq7} up to 1st order terms in \( |{\Delta }|=a \)
and obtain the local current density as a sum of two contributions,
field-driven and diffusive:\begin{equation}
\label{eq11}
{\mathbf{j}(r)}={\mathbf{j}}_{field}({\mathbf{r}})+{\mathbf{j}}_{dif}({\mathbf{r}})=G\left[ \rho ({\mathbf{r}})\right] {\mathbf{F}(r)}-eD\left[ \rho ({\mathbf{r}})\right] {\nabla }_{2}\sigma ({\mathbf{r}}),
\end{equation}
 where the effective conductivity \( G \) and diffusion coefficient
\( D \) are functions of local charge carrier density \( \rho  \):\begin{equation}
\label{eq12}
G\left( \rho \right) =\frac{4e^{2}\left( 1-\rho \right) }{a}\left| \rho \varphi ^{\prime }\left( 0\right) +2\left( 1-\rho \right) \varphi ^{\prime }\left( E_{c}\right) +\beta \varphi \left( E_{c}\right) \right| ,
\end{equation}
\begin{equation}
\label{eq13}
D\left( \rho \right) =\frac{\left( 1-\rho \right) \left[ \rho \varphi \left( 0\right) -2\left( 1-\rho \right) \varphi \left( E_{c}\right) \right] }{a\left( \rho -\sigma ^{2}\right) }.
\end{equation}
 Note that \( \rho -\sigma ^{2} \) in Eq. \ref{eq13} is nothing
but the mean square dispersion of charge density, and it is expressed
through \( \rho  \) by Eq. \ref{eq10}. In view of this equation,
we can also consider \( G \) and \( D \) as \emph{even} functions
of local charge density \( \sigma  \), and just this dependence will
be mostly used below. Also \( G \) and \( D \) depend on temperature,
through the functions \( \varphi  \), \( \varphi ^{\prime } \).
The system of Eqs. \ref{eq10}-\ref{eq13}, together with Eq. \ref{eq4},
is closed and self-consistent, defining the distributions of \( \sigma ({\mathbf{r}}) \)
and \( \rho ({\mathbf{r}}) \) at given \( {\mathbf{j}(r)} \).

Substituting Eq. \ref{eq11} into Eq. \ref{eq8} and considering the
low temperature limit when the diffusion coefficient \( D \) is practically
independent of \( \sigma  \), we transform the kinetic equation to
the form:

\begin{equation}
\label{eq14}
\dot{\sigma }-a^{2}bD\triangle _{2}\sigma =-\frac{a^{2}b}{e}{\nabla }_{2}\left[ G(\sigma ){\mathbf{F}}\right] \cdot 
\end{equation}
 where the l.h.s. has an aspect of common 2D diffusion equation and
the r.h.s. plays the role of source (or drain) function. For the FA
case, with the local field \( {\mathbf{F}} \) defined by Eq. \ref{eq4}
and the only relevant coordinate being \( x \), along the current
\( {\mathbf{j}(r)}=\rm const \) (Fig. 3), Eq. \ref{eq14} under steady
state condition (\( \dot{\sigma }=0 \)) leads to a (non-linear) integral
equation for \( \sigma (x) \):\begin{equation}
\label{eq14a}
\sigma (x)=\frac{1}{eD}\int _{0}^{x}\left\{ G\left[ \sigma (x^{\prime })\right] \left[ {\mathbf{F}}_{ext}+\frac{e}{\varepsilon _{eff}a^{2}}\int ^{L/2}_{-L/2}\frac{\sigma (x^{\prime \prime })}{x^{\prime }-x^{\prime \prime }}dx^{\prime \prime }\right] -j\right\} dx^{\prime }.
\end{equation}
Numeric analysis of Eq. \ref{eq14a} shows that there is no other
its solution but the trivial one: \( \sigma ({\mathbf{r}})\equiv 0 \).
Hence there is no diffusive contribution to the current, and the steady
state of FA in out-of-equilibrium conditions has an ohmic resistivity:\begin{equation}
\label{eq15}
r=\frac{1}{G\left( \rho _{0}\right) }.
\end{equation}
 With the parameter values \( a\sim 5 \) nm, \( \omega \sim 10^{12} \)
s\( ^{-1} \), \( N_{{\textrm{F}}}\sim 1 \) eV\( ^{-1} \), \( E_{c}\sim 10 \)
meV, \( T\sim 12 \) K, used in Eqs. \ref{eq12}, \ref{eq10}, we
have \( r\sim 50 \) \( \Omega  \) cm, which is in a reasonable agreement
with the measured resistance \( R\sim 50\div 100 \) M\( \Omega  \)
in a 10 layer granular sample of \( \sim 1\times 1 \) cm area \cite{Kak}.

\section{\label{ca}Steady state conduction in CA}

The kinetics in CA includes, besides the processes 1) to 3) of Sec.
3, also 4 additional microscopic processes between \( \mathbf{n} \)th
granule and the electrode (Fig. 6) which are just responsible for
variations of total charge \( Q \) by \( \pm 1 \). The respective
rates \( q^{\left( i\right) } \), \( i=4 \),\ldots{},\( 7 \),
are also dependent on the charging state \( (\sigma ,\rho ) \) of
the relevant granule and, using the same techniques that before for
FA, we obtain their mean values as:\begin{equation}
\label{eq16}
q^{\left( 4\right) }=\left( \rho +\sigma \right) \psi \left( -U-E_{c}^{\prime }\right) ,\quad q^{\left( 5\right) }=\left( \rho -\sigma \right) \psi \left( U-E_{c}^{\prime }\right) ,
\end{equation}
 \[
q^{\left( 6\right) }=\left( 1-\rho \right) \psi \left( U+E_{c}^{\prime }\right) ,\quad q^{\left( 7\right) }=\left( 1-\rho \right) \psi \left( -U+E_{c}^{\prime }\right) .\]
 Here the function \( \psi \left( E\right)  \) formally differs from
\( \varphi \left( E\right)  \) only by changing the prefactor: \( \omega \rightarrow \omega ^{\prime }=\omega _{a}{\textrm{e}}^{-2\chi b^{\prime }}\ll \omega  \),
but the arguments of these functions in Eq. \ref{eq16} include other
characteristic energies. Thus, \( U=eb^{\prime }F_{c} \) is generated
by the electric field \( F_{c} \) between granule and contact surface.
This field is always normal to the surface (see Fig. 7) and its value
is defined by the local charge density \( \sigma  \). At least, the
charging energy \( E_{c}^{\prime } \) for a granule under the contact
can be somewhat reduced (e.g., by \( \sim 1/2 \)) compared to \( E_{c} \).
Then the kinetic equations in CA present a generalization of Eqs.
\ref{eq5},\ref{eq5a}, as follows:\begin{equation}
\label{eq17}
\dot{\sigma }({\mathbf{r}})=\sum _{{\Delta }}\left[ q^{\left( 1\right) }({\mathbf{r},\Delta })-q^{\left( 1\right) }({\mathbf{r}+\Delta ,-\Delta })-p^{\left( 2\right) }({\mathbf{r},\Delta })+\right. 
\end{equation}
 \[
\left. +p^{\left( 2\right) }({\mathbf{r}+\Delta ,-\Delta })-p^{\left( 3\right) }({\mathbf{r}},{\Delta })-q^{\left( 4\right) }({\mathbf{r}})+q^{\left( 5\right) }({\mathbf{r}})+q^{\left( 6\right) }({\mathbf{r}})-q^{\left( 7\right) }({\mathbf{r}})\right] ,\]

\[
\dot{\rho }({\mathbf{r}})=\sum _{{\Delta }}\left[ q^{\left( 1\right) }({\mathbf{r},\Delta })+q^{\left( 1\right) }({\mathbf{r}+\Delta ,-\Delta })-q^{\left( 2\right) }({\mathbf{r},\Delta })+\right. \]
 \[
\left. +q^{\left( 2\right) }({\mathbf{r}+\Delta ,-\Delta })-q^{\left( 3\right) }({\mathbf{r}},{\Delta })-q^{\left( 4\right) }({\mathbf{r}})-q^{\left( 5\right) }({\mathbf{r}})+q^{\left( 6\right) }({\mathbf{r}})+q^{\left( 7\right) }({\mathbf{r}})\right] .\]
 The additional terms, by the {}``normal{}'' processes 4 to 7, are
responsible for appearance of a \emph{normal} component of current
density:\begin{equation}
\label{eq18}
{\mathbf{j}}_{z}\left( {\mathbf{r}}\right) =\frac{e{\mathbf{b}}}{a^{2}b}\left[ q^{\left( 4\right) }({\mathbf{r}})-q^{\left( 5\right) }({\mathbf{r}})-q^{\left( 6\right) }({\mathbf{r}})+q^{\left( 7\right) }({\mathbf{r}})\right] ,
\end{equation}
 besides the planar component, still given by Eq. \ref{eq7}. But
even more important difference from the FA case consists in the fact
that the Coulomb field here is formed by a \emph{double layer} of
charges, those of granules themselves and their images in the metallic
electrode (Fig. 7). Summing the contibutions from all the charged
granules and their images (except for the image of \( \mathbf{n} \)th
granule itself, already included in the energy \( E_{c}^{\prime } \)),
we arrive at the expression for the above mentioned field \( F_{c} \)
at the contact surface as a \emph{local} function of the charge density
\( \sigma ({\mathbf{r}}) \):\begin{equation}
\label{19}
F_{c}({\mathbf{r}})=F^{Coul}({\mathbf{r}},z=b^{\prime })=-\frac{4\pi e\sigma ({\mathbf{r}})}{\varepsilon _{eff}a^{2}},
\end{equation}
 (instead of integral relations, Eqs. \ref{eq3},\ref{eq4}, in FA).
Then, the planar component of the field by charged granules \( {\mathbf{F}}_{pl}({\mathbf{r}})={\mathbf{F}}^{Coul}({\mathbf{r}},z=0) \)
is determined by the above defined normal field \( F_{c} \) through
the relation \( {\mathbf{F}}_{pl}({\mathbf{r}})=b^{\prime }{\nabla }_{2}F_{c}({\mathbf{r}}) \).
The density of planar current is \( {\mathbf{j}}_{pl}({\mathbf{r}})=G{\mathbf{F}}_{pl}({\mathbf{r}})-eD{\nabla }_{2}\sigma ({\mathbf{r}}) \),
accordingly to Eq. \ref{eq11}, that is both field-driven and diffusive
contributions into \( {\mathbf{j}}_{pl} \) are present here and both
they are proportional to the gradient of \( \sigma ({\mathbf{r}}) \).
In the low temperature limit, this proportionality is given by:\begin{equation}
\label{eq20}
{\mathbf{j}}_{pl}({\mathbf{r}})\approx -\left\{ \frac{8\pi e^{3}\omega N_{{\textrm{F}}}b^{\prime }}{\varepsilon _{eff}a^{3}}g\left[ \sigma ({\mathbf{r}})\right] +\frac{e\omega N_{{\textrm{F}}}k_{{\textrm{B}}}T}{a}\right\} {\nabla }_{2}\sigma ({\mathbf{r}}).
\end{equation}
 Note that presence of a non-linear function:\begin{equation}
\label{eq21}
g\left( \sigma \right) =\sqrt{\rho _{0}^{2}+\sigma ^{2}}-\rho _{0}^{2}-\sigma ^{2},
\end{equation}
 defines a \emph{non-ohmic} conduction in CA. In fact, this function
should be given by Eq. \ref{eq21} up to maximum possible charge density
\( |\sigma _{max}|=\sqrt{1-\rho _{0}^{2}} \), turning zero for \( |\sigma |>|\sigma _{max}| \)
(note that the latter restriction just corresponds to our initial
limitation to the single charged states, see Sec. \ref{char}). In
the same limit of low temperatures, the normal current density is
obtained from Eqs. \ref{eq16},\ref{eq18} as \( {\mathbf{j}}_{z}({\mathbf{r}})=G_{z}{\mathbf{F}}_{c}({\mathbf{r}}) \)
where \( G_{z}\approx \omega ^{\prime }N_{{\textrm{F}}}E_{c}^{\prime }\varepsilon _{eff}/4\pi  \).
Finally, the kinetic equation in this case is obtained, in analogy
with Eq. \ref{eq8}, as:\begin{equation}
\label{eq22}
\dot{\sigma }\left( {\mathbf{r}}\right) =-\frac{a^{2}b}{e}{\nabla }_{2}\cdot {\mathbf{j}}_{pl}({\mathbf{r}})+\frac{a^{2}}{e}j_{z}({\mathbf{r}}).
\end{equation}
 This equation permits to describe the steady state conduction as
well as various time dependent processes. The first important conclusion
is that steady state conduction in CA turns only possible at non-zero
charge density gradient, that is, \emph{necessarily} involving charge
accumulation, in contrast to the above considered situation in FA.

Let us begin from the steady state regime which is simpler, in order
to use the obtained results later for the analysis of a more involved
case when an explicit temporal dependence of charge density is included
in Eq. \ref{eq22}.

We choose the CA geometry in the form of a rectangular stripe of planar
dimensions \( L\times L^{\prime } \), along and across the current
respectively. In neglect of relatively small effects of current non-uniformity
along the lateral boundaries, the only relevant coordinate for the
problem is longitudinal, \( x \) (Fig. 8). In the steady state regime,
the temporal derivative \( \dot{\sigma } \) in Eq. \ref{eq22} is
zero and the total current \( I=const \), defined by the action of
external source. Then, using Eq. \ref{eq21}, we arrive at a non-linear
2nd order equation for charge density:\begin{equation}
\label{eq23}
\frac{d}{dx}\left\{ g\left[ \sigma \left( x\right) \right] +\tau \right\} \frac{d\sigma \left( x\right) }{dx}-k^{2}\sigma \left( x\right) =0.
\end{equation}
 The parameters in Eq. \ref{eq23} are: \( k^{2}=(\omega ^{\prime }E_{c}^{\prime })/(ab\omega k_{{\textrm{B}}}T_{1}) \)
and \( \tau =T/T_{1} \), where \( T \) is the actual temperature
and \( T_{1}=8\pi e^{2}b^{\prime }/a^{2}b\varepsilon _{eff} \). To
define completely its solution, the following boundary conditions
are imposed:\begin{equation}
\label{eq24}
\left. \frac{d\sigma \left( x\right) }{dx}\right| _{x=0}=\frac{k^{2}b^{\prime }\sigma \left( x=0\right) }{g\left[ \sigma \left( 0\right) \right] +\tau }
\end{equation}
\begin{equation}
\label{eq25}
\left. \frac{d\sigma \left( x\right) }{dx}\right| _{x=L}=\frac{a}{Le\omega bN_{{\textrm{F}}}k_{{\textrm{B}}}T_{1}}\frac{I}{g\left[ \sigma \left( x=L\right) \right] +\tau }
\end{equation}
The condition \ref{eq24} corresponds to the fact that the longitudinal
current \( j_{x} \) at the initial point of CA (the leftmost in Fig.
8) is fully supplied by the normal current \( j_{y} \) entering from
the contact to granular sample, and the condition \ref{eq25} corresponds
to current continuity at passing from CA (of length \( L \) along
the current) to FA.

Let us discuss the solution of Eq. \ref{eq23} qualitatively. Generally,
to fulfill the conditions, Eqs. \ref{eq24},\ref{eq25}, one needs
a quite subtle balance to be maintained between the charge density
and its derivatives at both ends of CA. But the situation is radically
simplified when the length \( L \) is much greater than the characteristic
decay length for charge and current density: \( kL\gg 1 \). In this
case, the relevant coordinate is \( \xi =L-x \), so that the boundary
condition \ref{eq24} corresponds to \( \xi =L\rightarrow \infty  \),
when both its left and right hand side turn zeros:\begin{equation}
\label{eq26}
\left. \sigma \left( \xi \right) \right| _{\xi \rightarrow \infty }=0,\qquad \left. \frac{d\sigma \left( \xi \right) }{d\xi }\right| _{\xi \rightarrow \infty }=0.
\end{equation}
 The numeric solution shows that, for any initial (with respect to
\( \xi  \), that is related to \( x=L \), Eq. \ref{eq25}) value
of charge density \( \sigma \left( \xi =0\right) =\sigma _{0} \),
there is a \emph{unique} initial value of its derivative \( d\sigma \left( \xi \right) /d\xi |_{\xi =0}=D(\sigma _{0}) \)
which just assures the limits \ref{eq26}, while for \( d\sigma \left( \xi \right) /d\xi |_{\xi =0}>D(\sigma _{0}) \)
the asymptotic value diverges as \( \sigma \left( \xi \rightarrow \infty \right) \rightarrow \infty  \),
and for \( d\sigma \left( \xi \right) /d\xi |_{\xi =0}<D(\sigma _{0}) \)
it diverges as \( \sigma \left( \xi \rightarrow \infty \right) \rightarrow -\infty  \).
Then, using the boundary condition \ref{eq25} and taking into account
the relation \( V=V_{0}\sigma _{0} \) following from Eq. \ref{19}
with \( V_{0}=4\pi eb^{\prime }/(\varepsilon _{eff}a^{2}) \), we
conclude that the function \( D(\sigma _{0}) \) generates the \emph{I}-\emph{V}
characteristics for CA:\begin{equation}
\label{eq27}
I=I_{1}b^{\prime }D\left( V/V_{0}\right) \left[ g\left( V/V_{0}\right) +\tau \right] 
\end{equation}
 where \( I_{1}=e\omega N_{{\textrm{F}}}k_{{\textrm{B}}}T_{1} \).

A more detailed analysis of Eq. \ref{eq23} is presented in Appendix.
In particular, for the weak current regime (regime I) when \( \sigma _{0}\ll \sigma _{1}=\sqrt{32\rho _{0}(\rho _{0}+\tau )}\ll 1 \),
so that \( g(\sigma )\approx \rho _{0}+\sigma ^{2}/2\rho _{0} \)
along whole CA, Eq. \ref{eq23} admits an approximate analytic solution:\begin{equation}
\label{eq28}
\sigma \left( \xi \right) =\sigma _{0}{\textrm{e}}^{-\lambda \xi }\left[ 1+6\left( \frac{\sigma _{0}}{\sigma _{1}}\right) ^{2}\left( 1-{\textrm{e}}^{-2\lambda \xi }\right) \right] ,
\end{equation}
 with the exponential decay index \( \lambda =k/\sqrt{\rho _{0}+\tau } \).
This results in the explicit \emph{I}-\emph{V} characteristics for
regime I:\begin{equation}
\label{eq29}
I=\frac{V}{R_{0}\left( \tau \right) }\left[ 1+\left( \frac{V}{V_{1}}\right) ^{2}\right] ,
\end{equation}
 presented by the line 1 in Fig. 9. For \( V<V_{1}=\sigma _{1}V_{0} \),
Eq. \ref{eq29} describes the initial ohmic resistence (dependent
on temperature \( \tau  \)):\begin{equation}
\label{eq30}
R_{0}\left( \tau \right) =\frac{V_{0}}{I_{1}kb^{\prime }\sqrt{\rho _{0}\left( \tau \right) +\tau }},
\end{equation}
 which turns non-ohmic for \( V\sim V_{1} \). But at so high voltages
another conduction regime already applies (called regime II), where
\( \sigma _{1}\ll \sigma _{0}\ll 1 \) and one has \( g(\sigma )\approx \sigma  \)
(see Eq. \ref{eq21}). Following the same reasoning as for the regime
I, we obtain a non-linear \emph{I}-\emph{V} characteristics for regime
II:\begin{equation}
\label{31}
I\approx \frac{I_{1}kb^{\prime }}{\sqrt{3}}\left( \frac{V}{V_{0}}+\tau \right) ^{3/2},
\end{equation}
 presented by the line 2 in Fig. 9. This law is weaker temperature
dependent than Eq. \ref{eq29}, which relates to the fact that the
conductance in regime II is mainly due to dynamical accumulation of
charge and not to thermic excitation of charge carriers. However,
such dependence can be more pronounced if multiple charging states
are engaged, as may be the case in real granular layers with a certain
statistical distribution of granule sizes present.

At least, for even stronger currents, when already \( \sigma _{0}\sim 1 \),
the solutions of Eq. \ref{eq23} can be obtained numerically, following
the above discussed procedure of adjustment of the derivative \( D(\sigma _{0}) \)
to a given \( \sigma _{0} \). Such solutions define the line 3 in
Fig. 9, which generally agrees with the lines 1 and 2, but its asymptotics
turns to \( I\sim V^{5/4} \).

A simple and important exact relation for the total accumulated charge
\( Q \) in CA is obtained from the direct integration of Eq. \ref{eq23}:\begin{equation}
\label{eq32}
Q=\tau I,
\end{equation}
 where the parameter \( \tau =1/\psi \left( -E_{c}^{\prime }\right)  \)
has a role of characteristic relaxation time in non-stationary processes,
as will be seen below. Assuming its value \( \tau \sim 1 \) s (comparable
with the experimental observations \cite{Kak}), together with the
above used values of \( \omega  \) and \( T_{1} \), we conclude
that the characteristic length scale \( \lambda ^{-1} \) for solutions
of Eq. \ref{eq23} can reach up to \( \sim 10^{4}a\sim 0.1 \) mm,
which is a reasonable scale for a charge distribution beneath the
contacts.

\section{\label{tdep}Time-dependent processes}

Now we can extend our analysis also to a non-stationary situation.
Actually, let us consider the simplest example of this type, that
of interrupting the circuit of Fig. 3 at the moment \( t_{0}=0 \),
after an infinitely long steady state with current \( I \). Beginning
from this moment, the constant value is not the total current, as
was the case in Sec. \ref{mf}, but the total charge in the overall
system composed by the granular sample itself and electrodes (in fact,
this total value is zero and it corresponds to the uniform distribution
\( \sigma ({\mathbf{r}},\infty )\equiv 0 \) in the asymptotic limit
\( t\rightarrow \infty  \)). We use the same 1D geometry and, accordingly
to the above mentioned absence of charges in FA, limit ourselves only
to consideration of CA. The initial distribution \( \sigma (x) \),
corresponding to the steady state, will degrade at \( t>t_{0} \),
at the expense of recombination of charges accumulated in granules
beneath the contact with the polarization charges on the contact surface
(the {}``image{}'' charges). This process is described by Eq. \ref{eq22},
subject to the initial condition \( \sigma (x,t=t_{0})\equiv 0 \)
and the boundary conditions:\begin{equation}
\label{eq33}
\dot{\sigma }\left( L,t\right) =-\frac{\sigma \left( L,t\right) }{\tau }+\frac{g\left[ \sigma \left( x,t\right) \right] +\tau }{\tau bk^{2}}\left. \frac{d\sigma \left( x,t\right) }{dx}\right| _{x=L}+\frac{I-C\dot{V}}{eL^{\prime }b},
\end{equation}
\begin{equation}
\label{eq34}
\dot{\sigma }\left( 0,t\right) =-\frac{\sigma \left( 0,t\right) }{\tau }-\frac{g\left[ \sigma \left( x,t\right) \right] +\tau }{\tau bk^{2}}\left. \frac{d\sigma \left( x,t\right) }{dx}\right| _{x=0},
\end{equation}
 where \( V \) and \( C \) are respectively the voltage and capacitance
between the contacts. In similarity with the above considered steady
state case, the non-stationary Eq. \ref{eq22} with the boundary conditions,
Eqs. \ref{eq33},\ref{eq34}, can be solved numerically, in fact presenting
an exponential decay with characteristic time \( \sim \tau  \) (the
corresponding plot not shown here).

But a more transparent description of dynamical processes in the considered
system is achieved with using the equivalent circuit (Fig. 10), following
from the previous analisis. Here \( R \) represents the resistence
of FA, which is ohmic (due to the absence of charges) but strongly
temperature dependent. The two resistances \( R^{\prime } \) associated
to each CA are already non-ohmic, as seen from Eqs. \ref{eq20},\ref{eq21},
due to the accumulated charge \( Q \). The proportionality between
\( Q \) and the current \( I_{{\textrm{CA}}} \) through CA, Eq.
\ref{eq32}, together with the simplifying assumption of uniform charge
distribution over the rectangular CA, permits to express this non-ohmic
behavior through the function \( g \) given by Eq. \ref{eq21}: \begin{equation}
\label{eq35}
R^{\prime }\left( I_{{\textrm{CA}}}\right) =\frac{R_{0}}{g\left( I_{{\textrm{CA}}}/I_{0}\right) }.
\end{equation}
 Here the resistance scale \( R_{0}=aL/(4bL^{\prime }e^{2}\varphi ^{\prime }\left( 0\right) ) \)
can be fixed by a direct comparison with the experimentally measured
steady state resistance of the whole circuit, and the current scale
\( I_{0}=eLL^{\prime }/(\tau a^{2}) \) is about \( 1 \) \( \mu  \).A
for \( L\sim L^{\prime }\sim 1 \) cm and \( a\sim 5 \) nm. The respective
capacities have the orders of magnitude: \( C_{1}\sim L \) (capacity
between the electrodes through FA) and \( C_{2}\sim LL^{\prime }/b \)
(capacity between an electrode and adjacent granular layer), hence
\( C_{1}\ll C_{2} \). Then the temporal evolution of the system is
given by the equations for voltages \( V_{1} \) and \( V_{2} \)
on them:\begin{equation}
\label{eq36}
\dot{V}_{1}=\frac{2V_{2}-V_{1}}{RC},
\end{equation}
 \[
\dot{V}_{2}=\frac{V_{1}}{RC^{\prime }}-\left( \frac{2}{RC^{\prime }}+\frac{g\left( V_{2}\right) }{C^{\prime }}\right) V_{2},\]

The numeric solution to this equivalent system with initial conditions
\( V_{1}(0)=V \), \( V_{2}(0)=(V-IR_{1})/2 \), adequate to the considered
discharge regime, is presented in Fig. 11. It can be noted that the
character of decay is very similar to that obtained with the exact
kinetic equation, Eq. \ref{eq22}. Another intersting moment is the
presence of two {}``relaxation times{}'' seen in a log plot (lower
panel of Fig. 11). The initial discharge phase, with the strongest
current through CA and the lowest resistance \( R^{\prime } \) (by
Eq. \ref{eq35}), has a very fast decay whereas the final phase has
a much slower relaxation rate, due to the growth of \( R^{\prime } \).
This behavior also agrees with the experimental observations at switching
off the current through a granular layer \cite{Kak}, revealing two
strongly different relaxation times, the faster one being temperature
dependent.

The relatively simple non-linear system, Eqs. \ref{eq36}, can be
also used for treating more complicate transient processes.

\section{\label{ap}Appendix}

Let us consider the equation:\begin{equation}
\label{eq37}
\frac{d}{d\xi }\left[ g\left( \sigma \right) +\tau \right] \frac{d\sigma }{d\xi }-k^{2}\sigma =0
\end{equation}
 with certain boundary conditions \( \sigma (0)=\sigma _{0} \), \( \sigma ^{\prime }(0)=\sigma ^{\prime }_{0} \),
resulting from Eqs. \ref{eq24},\ref{eq25}. For a rather general
function \( g\left( \sigma \right)  \) we can define the function\begin{equation}
\label{eq38}
f\left( \sigma \right) =\int _{0}^{\sigma }g\left( \sigma ^{\prime }\right) d\sigma ^{\prime },
\end{equation}
 then Eq. \ref{eq37} presents itself as:\begin{equation}
\label{eq39}
\frac{d^{2}F\left( \xi \right) }{d\xi ^{2}}=k^{2}\sigma \left( \xi \right) ,
\end{equation}
 where \( F\left( \xi \right) \equiv f[\sigma (\xi )]+\tau \sigma (\xi ) \).
Considered irrespectively of \( \xi  \): \begin{equation}
\label{eq39a}
f(\sigma )+\tau \sigma =F,
\end{equation}
 this equation also defines \( \sigma  \) as a certain function of
\( F \): \( \sigma =\sigma (F) \). Hence it is possible to construct
the following function:\begin{equation}
\label{eq40}
\varphi \left( F\right) =2\int _{0}^{F}\sigma (F^{\prime })dF^{\prime }.
\end{equation}
 Now, multiplying Eq. \ref{eq39} by \( 2dF/d\xi  \), we arrive at
the equation:\begin{equation}
\label{eq41}
\frac{d}{d\xi }\left( \frac{dF}{d\xi }\right) ^{2}=k^{2}\frac{d\varphi }{d\xi },
\end{equation}
 with \( \varphi (\xi )\equiv \varphi [F(\xi )] \). Integrating Eq.
\ref{eq41} in \( \xi  \), we obtain a 1st order separable equation
for \( F(\xi ) \):\begin{equation}
\label{eq42}
\frac{dF}{d\xi }=\pm k\sqrt{\varphi (F)}.
\end{equation}
 We expect the function \( F \) to decrease at going from \( \xi =0 \)
into depth of CA, hence choose the negative sign on r.h.s. of Eq.
\ref{eq42} and obtain its explicit solution as\begin{equation}
\label{eq43}
\int _{F\left( \xi \right) }^{F_{0}}\frac{dF^{\prime }}{\sqrt{\varphi (F^{\prime })}}=k\xi 
\end{equation}
 with \( F_{0}=f(\sigma _{0})+\tau \sigma _{0} \). Finally, the sought
solution for \( \sigma (\xi )=\sigma [F(\xi )] \) results from substitution
of the function \( F(\xi ) \), given implicitly by Eq. \ref{eq43},
into \( \sigma (F) \) defined by Eq. \ref{eq39a}. Consider some
particular realizations of the above scheme.

For the function \( g(\sigma ) \) given by Eq. \ref{eq21}, we have
the explicit integral, Eq. \ref{eq38}, in the form:\begin{equation}
\label{eq44}
F\left( \sigma \right) =f\left( \sigma \right) +\tau \sigma =\left( \tau +\frac{\sqrt{\rho _{0}^{2}+\sigma ^{2}}}{2}-\rho _{0}^{2}-\frac{\sigma ^{2}}{3}\right) \sigma +\rho _{0}^{2}\ln \sqrt{\frac{\sigma +\sqrt{\rho _{0}^{2}+\sigma ^{2}}}{\rho _{0}}}.
\end{equation}

In the case \( \sigma \ll \rho _{0}\ll 1 \) (regime I), Eq. \ref{eq44}
is approximated as:\begin{equation}
\label{eq45}
F\approx \left( \rho _{0}+\tau \right) \sigma +\frac{\sigma ^{3}}{6\rho _{0}}
\end{equation}
 hence \( \sigma (F) \) corresponds to a real root of the cubic equation,
Eq. \ref{eq45}, and in the same approximation of regime I it is given
by:\begin{equation}
\label{eq46}
\sigma \left( F\right) \approx \frac{F}{\rho _{0}+\tau }\left( 1-\frac{8F^{2}}{\sigma _{1}^{2}}\right) ,
\end{equation}
 with \( \sigma _{1}=4\sqrt{\rho _{0}(\rho _{0}+\tau )^{3}} \). Using
this form in Eq. \ref{eq40}, we obtain:\begin{equation}
\label{eq47}
\varphi \left( F\right) \approx \frac{F^{2}}{\rho _{0}+\tau }\left( 1-\frac{4F^{2}}{\sigma _{1}^{2}}\right) ,
\end{equation}
 and then substituting into Eq. \ref{eq43}:\begin{equation}
\label{eq48}
\ln \frac{\left[ 1+\sqrt{1-\left( 2F/\sigma _{1}\right) ^{2}}\right] F_{0}}{\left[ 1+\sqrt{1-\left( 2F_{0}/\sigma _{1}\right) ^{2}}\right] F}=\lambda \xi .
\end{equation}
 Inverting this relation, we define an explicit solution for \( F(\xi ) \):\begin{equation}
\label{eq49}
F\left( \xi \right) \approx F_{0}{\textrm{e}}^{-\lambda \xi }\left[ 1+\frac{F_{0}^{2}}{\sigma _{1}^{2}}\left( 1-{\textrm{e}}^{-2\lambda \xi }\right) \right] .
\end{equation}
 Finally, substituting Eq. \ref{eq49} into Eq. \ref{eq46}, we arrive
at the result of Eq. \ref{eq28} corresponding to Fig. 12.

For the regime II we have in a similar way:\begin{equation}
\label{eq50}
F\left( \sigma \right) \approx \sigma \left( \tau +\sigma /2\right) ,\qquad \sigma \left( F\right) \approx \sqrt{2F+\tau ^{2}}-\tau ,
\end{equation}
 \[
\varphi \left( F\right) \approx \frac{2}{3}\left[ \left( 2F+\tau ^{2}\right) ^{3/2}-\tau \left( 3F+\tau ^{2}\right) \right] ,\qquad F\left( \xi \right) \approx \left[ F_{0}^{1/4}-\lambda _{1}\xi +\frac{3\tau }{2^{5/4}\left( F_{0}^{1/4}-\lambda _{1}\xi \right) }\right] ^{4},\]
 with \( \lambda _{1}=2^{1/4}k/(4\sqrt{3}) \), obtaining the charge
density distribution (Fig. 13e):\begin{equation}
\label{eq51}
\sigma \left( \xi \right) \approx \left( \sqrt{\sigma _{0}+\tau }-\lambda _{1}\xi \right) ^{2}-\tau .
\end{equation}
 This function seems to turn zero already at \( \xi =(\sqrt{\sigma _{0}+\tau }-\sqrt{\tau })/\lambda _{1} \),
but in fact the fast parabolic decay by Eq. \ref{eq51} only extends
to \( \xi \sim \xi ^{*} \), such that \( \sigma (\xi ^{*})\sim \rho _{0} \),
and for \( \xi >\xi ^{*} \) the decay turns exponential, like Eq.
\ref{eq28}. The \emph{I}-\emph{V} characteristics, Eq. \ref{31},
follows directly from Eq. \ref{eq51}.

\end{document}